\documentclass{PoS}
      \usepackage{bm}

\title{Production of \bm{$c \bar c c \bar c$}\\
 in single- and double-parton scattering \\ 
in collinear and \bm{$k_t$}-factorization approaches}

\ShortTitle{Double charm production in single- and double-parton scattering}

\author{\speaker{Antoni Szczurek}%
        \thanks{This work was partially supported by the Polish National
          Science Centre Grant No.2014/15/B/ST2/02528(OPUS).}\\
       Institute of Nuclear Physics PAN, PL-31-342 Cracow, Poland\\
       and Rzesz\'ow University, PL-35-959 Rzesz\'ow, Poland\\
       E-mail: \email{antoni.szczurek@ifj.edu.pl}}

\author{Rafal Maciula\\%
       Institute of Nuclear Physics PAN, PL-31-342 Cracow, Poland\\
       E-mail: \email{rafal.maciula@ifj.edu.pl}}

\abstract{We discuss production of two pairs of $c \bar c$ in proton-proton
collisions at the LHC.
Both double-parton scattering (DPS) and single-parton scattering (SPS)
contributions are included in the analysis. Each step of DPS is
calculated within $k_t$-factorization approach which effectively
includes some next-to-leading order corrections.
The discussed mechanisms unavoidably lead to the production of pairs of mesons:
$D_i D_j$ (each containing $c$ quarks) or $\bar D_i \bar D_j$ 
(each containing $\bar c$ antiquarks). We calculate corresponding 
differential distribution for
$(D^0 D^0$ + $\bar D^0 \bar D^0)$ production. Within large theoretical 
uncertainties the predicted DPS cross section is
fairly similar to the cross section measured recently by 
the LHCb collaboration. We also present first results for the $2 \to 4$ 
single-parton scattering $g g \to c \bar c c \bar c$ subprocess for 
the first time fully within the $k_t$-factorization approach. 
In this calculation we have used an off-shell matrix element squared 
calculated using recently developed techniques.
The results are compared with our earlier result obtained within the 
collinear approach. Only slightly larger cross sections are obtained 
than in the case of the collinear approach but still 
the SPS mechanism contribution is much smaller than the DPS one.
Inclusion of transverse momenta of gluons entering the hard process
leads to a much stronger azimuthal decorrelation between $c c$ and
$\bar c \bar c$ than in the collinear-factorization approach.
A comparison to predictions of double parton scattering (DPS) results
and the LHCb data strongly suggests that the assumption of two
fully independent DPS ($g g \to c \bar c \otimes g g \to c \bar c$)
may be too approximate.}

\FullConference{The European Physical Society Conference on High Energy Physics\\
          22-29 July 2015\\
          Vienna, Austria}

\begin{document}

\section{Introduction}

The $p p \to c \bar{c} c \bar{c} X$ reaction has been recognized
recently to be a golden reaction
to study double-parton scattering (DPS) processes 
\cite{Luszczak:2011zp,Maciula:2013kd}.
The LHCb collaboration confirmed the theoretical predictions
and obtained a large cross section for production of two mesons,
both containing $c$ quarks or both containing $\bar c$ antiquarks 
\cite{Aaij:2012dz}.
The single-parton scattering (SPS) contribution was discussed in 
Refs.~\cite{Schafer:2012tf} and \cite{vanHameren:2014ava}.
In the first case \cite{Schafer:2012tf} a high-energy approximation 
was used neglecting some unimportant at high energies Feynman diagrams. 
Last year we have calculated the lowest-order SPS cross section(s)
including a complete set of Feynman diagrams \cite{vanHameren:2014ava} 
in the collinear-factorization approach.
The final result was only slightly different than that obtained
in the high-energy approximation.

Now we go one step further and calculate the SPS cross sections for
the $p p \to c \bar c c \bar c X$ reaction consistently in 
the $k_t$-factorization approach \cite{vanHameren:2015}.
In this theoretical framework a sizeable part of higher-order
corrections can be included and studies
of kinematical correlations are available. 
From the technical point of view this is a first calculation
within the $k_t$-factorization approach based on a $2 \to 4$ 
subprocesses with two off-shell initial-state partons (gluons).
The result is important in the context of studying DPS as 
the considered SPS mechanism constitutes an irreducible background, 
and its estimation is therefore crucial if deeper conclusions
concerning DPS can be drawn from measurements at the LHC.

A convenient formalism for the automation of the calculation of 
tree-level scattering amplitudes with off-shell gluons for arbitrary 
processes was introduced recently in Ref.~\cite{vanHameren:2012if}.
Off-shell gluons are replaced by eikonal quark-antiquark pairs, and 
the amplitude can be calculated with the help of standard local Feynman
rules, including the eikonal gluon-quark-antiquark vertex and 
the eikonal quark-antiquark propagator.
The well-known successful recursive methods to calculate 
tree-level amplitudes can directly be applied, including 
the ``on-shell'' recursion, or Britto-Cachazo-Feng-Witten recursion, 
as shown in Ref.~\cite{vanHameren:2014iua}.
The heuristic introduction of the formalism in 
Ref.~\cite{vanHameren:2012if} has be given solid ground 
in Ref.~\cite{Kotko:2014aba}.

\section{Formalism}

In leading-order (LO) collinear approximation the differential distributions 
for $c\bar{c}$ production depend e.g. on the rapidity of the quark, 
the rapidity of the antiquark and the transverse momentum of one of 
them (they are identical). In the next-to-leading order (NLO) 
collinear approach or in the $k_t$-factorization approach the situation 
is more complicated as there are more kinematical variables necessary 
to describe the kinematical situation.
In the $k_t$-factorization approach the differential cross section for
DPS production of $c \bar c c \bar c$ system, assuming factorization 
of the DPS model, can be written as: 
\begin{eqnarray}
\frac{d \sigma^{DPS}(p p \to c \bar c c \bar c X)}{d y_1 d y_2 d^2 p_{1,t} d^2 p_{2,t} 
d y_3 d y_4 d^2 p_{3,t} d^2 p_{4,t}} = \nonumber \;\;\;\;\;\;\;\;\;\;\;\;\;\;\;\;\;\;\;\;\;\;\;\;\;\;\;\;\;\;\;\;\;\;\;\;\;\;\;\;\;\;\;\;\;\;\;\;\;\;\;\;\;\;\;\;\;\;\;\;\;\;\;\;\;\;\; \\ 
\frac{1}{2 \sigma_{eff}} \cdot
\frac{d \sigma^{SPS}(p p \to c \bar c X_1)}{d y_1 d y_2 d^2 p_{1,t} d^2 p_{2,t}}
\cdot
\frac{d \sigma^{SPS}(p p \to c \bar c X_2)}{d y_3 d y_4 d^2 p_{3,t} d^2 p_{4,t}}.
\end{eqnarray}

When integrating over kinematical variables one obtains
\begin{equation}
\sigma^{DPS}(p p \to c \bar c c \bar c X) = \frac{1}{2 \sigma_{eff}}
\sigma^{SPS}(p p \to c \bar c X_1) \cdot \sigma^{SPS}(p p \to c \bar c X_2).
\label{basic_formula}
\end{equation}
These formulae assume that the two parton subprocesses are not
correlated.

Experimental data obtained at Tevatron \cite{Tevatron1,Tevatron2} and 
LHC \cite{Aaij:2011yc,Aaij:2012dz,Aad:2013bjm} 
provide an estimate of $\sigma_{eff}$ in the denominator of formula 
(\ref{basic_formula}). Phenomenological studies of $\sigma_{eff}$ are 
summarized e.g. in Ref.~\cite{Seymour} with the average 
value $\sigma_{eff} \approx$ 15 mb. 

Within the $k_t$-factorization approach the SPS cross section for 
$p p \to c \bar c c \bar c X$ reaction can be written as
\begin{equation}
d \sigma_{p p \to c \bar c c \bar c} =
\int d x_1 \frac{d^2 k_{1t}}{\pi} d x_2 \frac{d^2 k_{2t}}{\pi}
{\cal F}(x_1,k_{1t}^2,\mu^2) {\cal F}(x_2,k_{2t}^2,\mu^2)
d {\hat \sigma}_{gg \to c \bar c c \bar c}
\; .
\label{cs_formula}
\end{equation}
In the formula above ${\cal F}(x,k_t^2,\mu^2)$ are unintegrated
gluon distributions that depend on longitudinal momentum fraction $x$,
transverse momentum squared $k_t^2$ of the gluons entering the hard process,
and in general also on a (factorization) scale of the hard process $\mu^2$.
The elementary cross section in Eq.~(\ref{cs_formula}) can be written
somewhat formally as:
\begin{eqnarray}
d {\hat \sigma} &=&
\frac{d^3 p_1}{2 E_1 (2 \pi)^3} \frac{d^3 p_2}{2 E_2 (2 \pi)^3}
\frac{d^3 p_3}{2 E_3 (2 \pi)^3} \frac{d^3 p_4}{2 E_4 (2 \pi)^3}
(2 \pi)^4 \delta^{4}(p_1 + p_2 + p_3 + p_4 - k_1 - k_2) \nonumber \\
&&\times\frac{1}{\mathrm{flux}} \overline{|{\cal M}_{g^* g^* \to c \bar c c \bar c}(k_{1},k_{2})|^2}
\; ,
\label{elementary_cs}
\end{eqnarray}
where only dependence of the matrix element on four-vectors of gluons 
$k_1$ and $k_2$ is made explicit. In general all four-momenta associated
with partonic legs enter.
The matrix element takes into account that both gluons entering 
the hard process are off-shell with virtualities 
$k_1^2 = -k_{1t}^2$ and $k_2^2 = -k_{2t}^2$.
The matrix element squared is rather complicated
and explicit formula will be not given here.

\section{Results}

In this section we compare the new results of the $k_t$-factorization 
approach for SPS mechanism to those obtained by us in 
Ref.~\cite{vanHameren:2014ava} in the collinear-factorization approach.

\begin{figure}[!h]
\begin{minipage}{0.47\textwidth}
 \centerline{\includegraphics[width=1.0\textwidth]{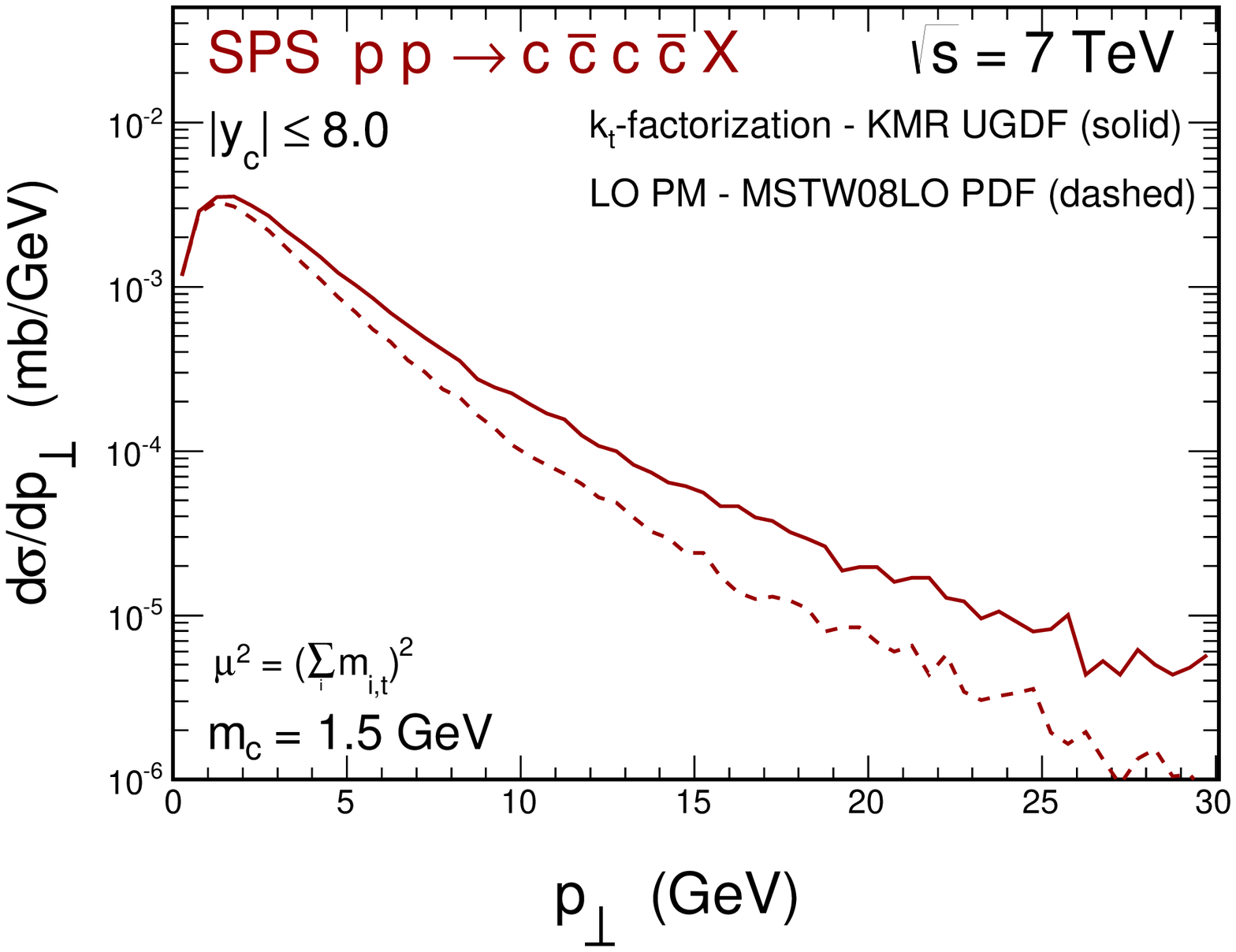}}
\end{minipage}
\hspace{0.5cm}
\begin{minipage}{0.47\textwidth}
 \centerline{\includegraphics[width=1.0\textwidth]{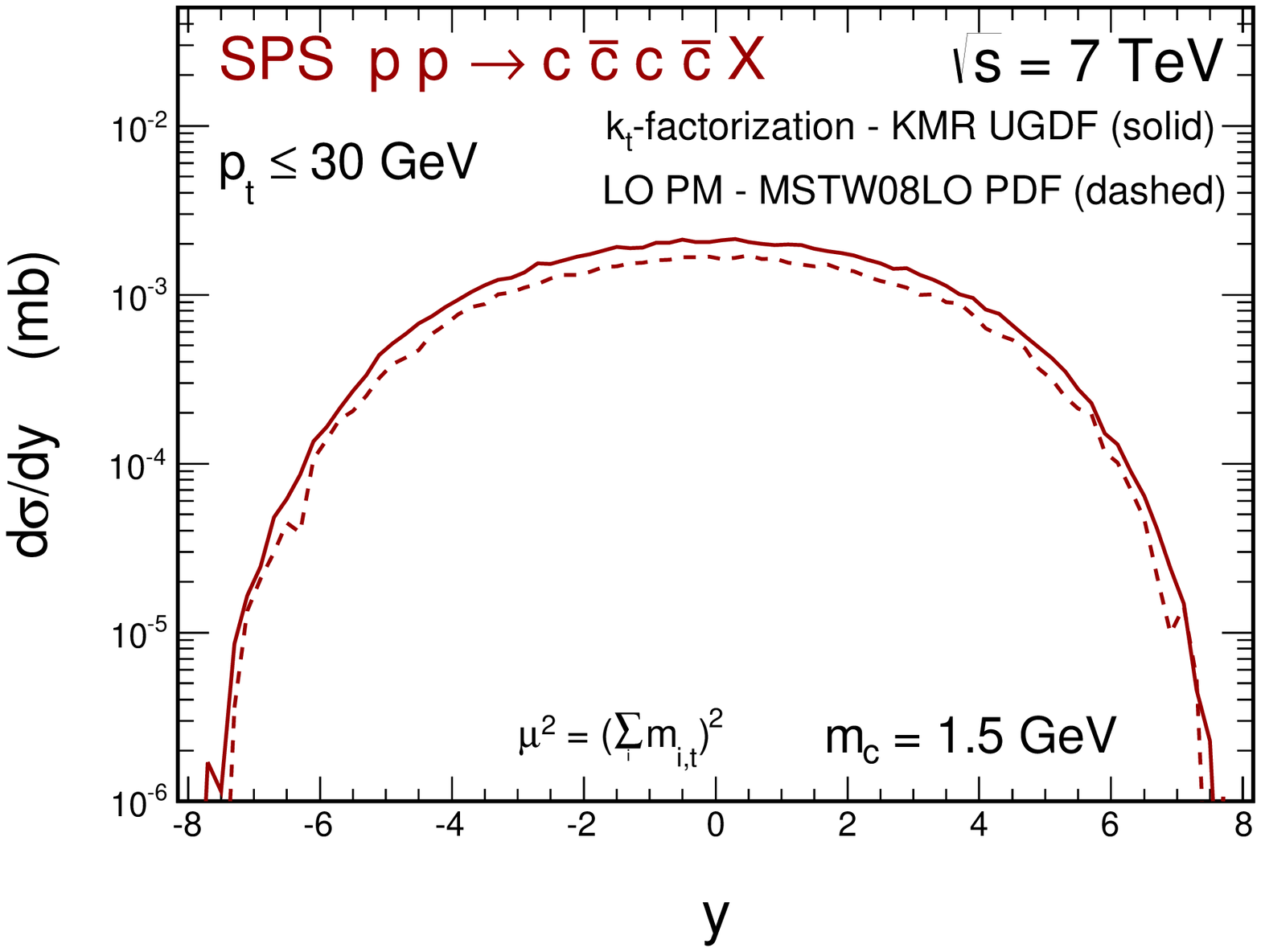}}
\end{minipage}
\caption{
\small Distributions in $c$ quark ($\bar c$ antiquark) transverse momentum
(left panel) and rapidity (right panel). The $k_t$-factorization result
(solid line) is compared with the collinear-factorization result (dashed
line).
}
 \label{fig:dsig_dpt_dy}
\end{figure}

In Fig.~\ref{fig:dsig_dpt_dy} we show standard single
particle distributions in charm quark/antiquark transverse momentum
(left panel) and its rapidity (right panel). We predict an enhancement
of the cross section at large transverse momenta of $c$ or $\bar{c}$ 
compared to the collinear-factorization approach.
The rapidity distributions in both approaches are rather similar
(see the left panel of the figure).

\begin{figure}[!h]
\begin{minipage}{0.47\textwidth}
 \centerline{\includegraphics[width=1.0\textwidth]{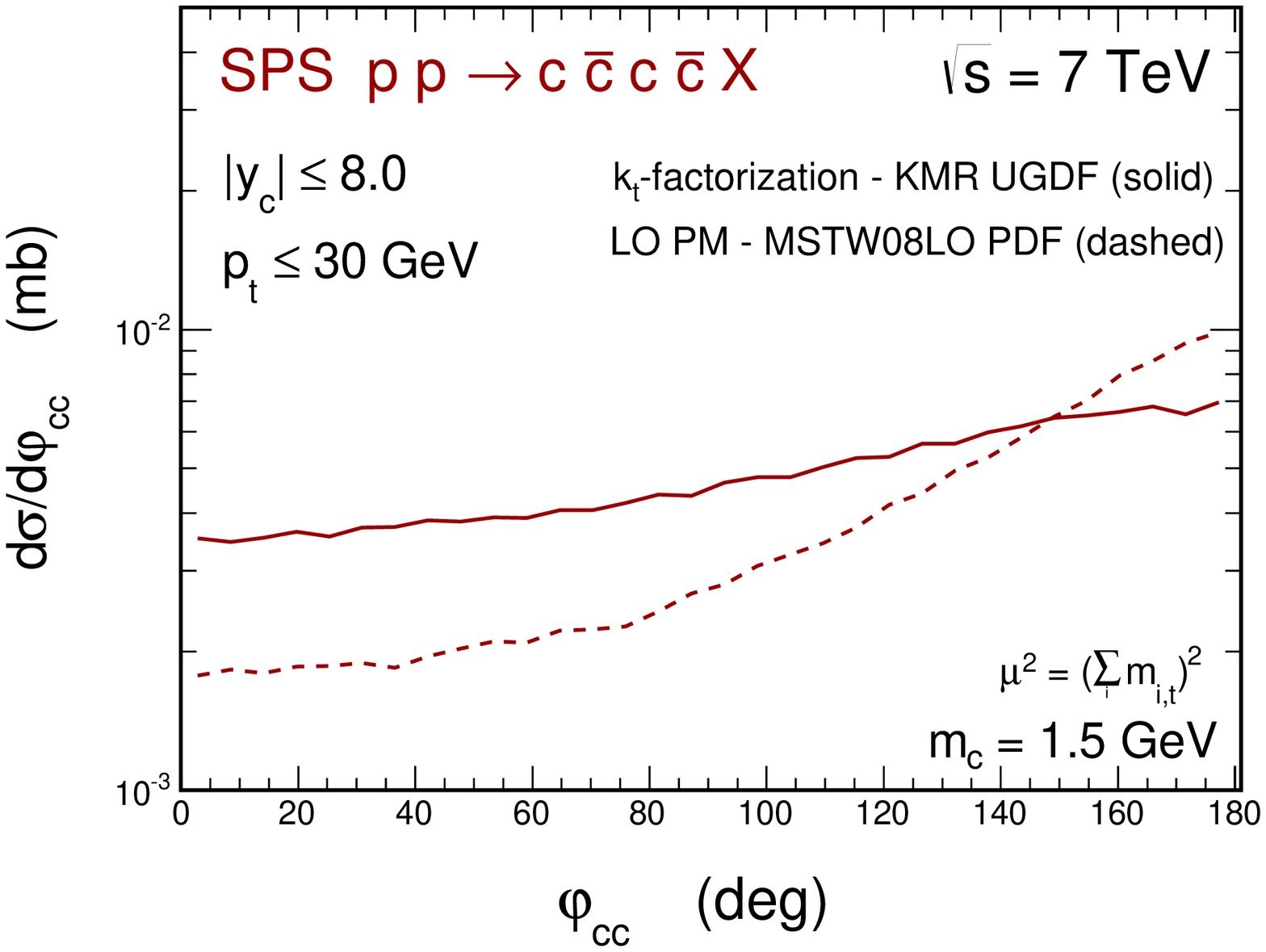}}
\end{minipage}
\hspace{0.5cm}
\begin{minipage}{0.47\textwidth}
 \centerline{\includegraphics[width=1.0\textwidth]{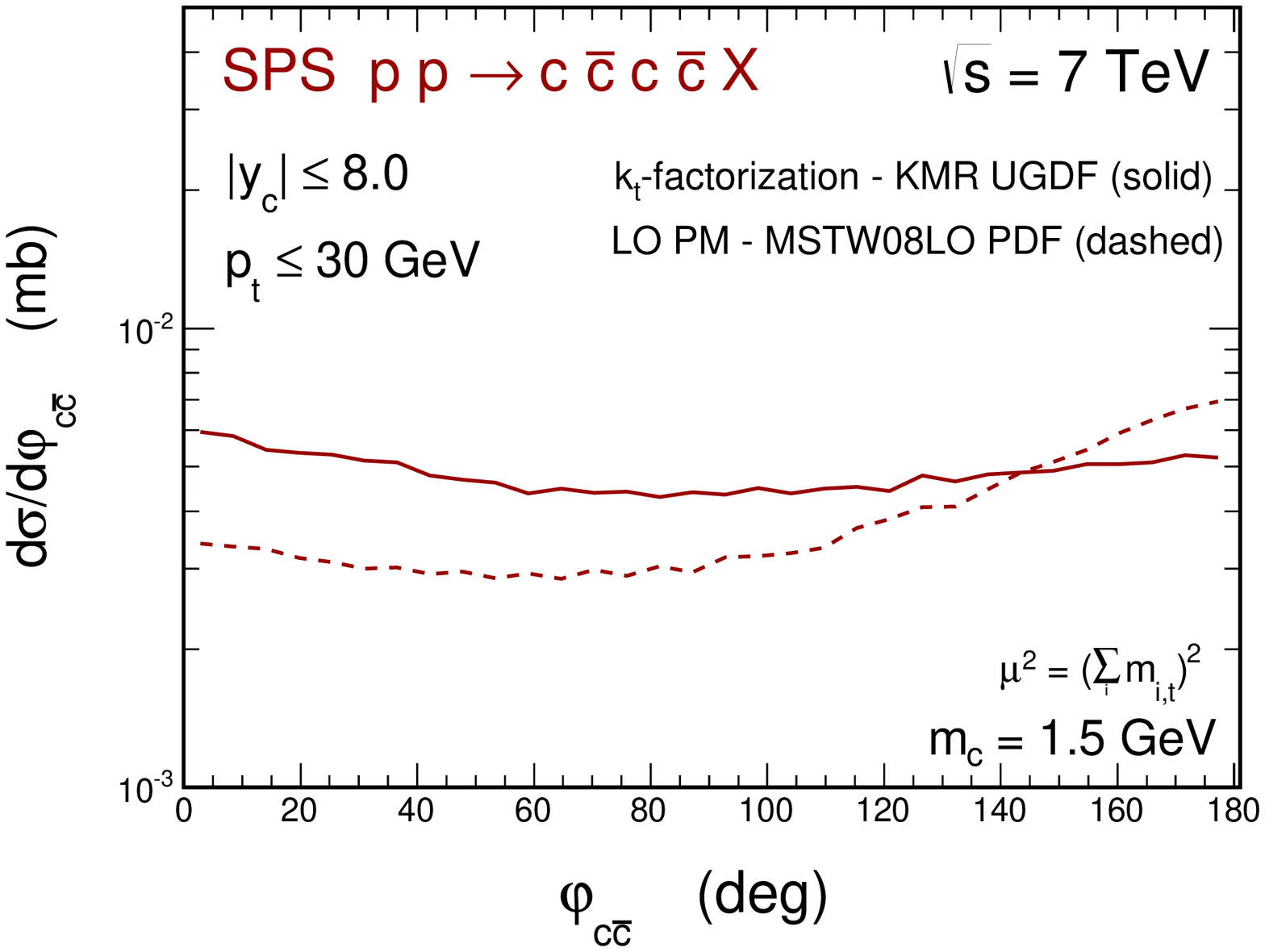}}
\end{minipage}

\caption{
\small Azimuthal angle correlations between two $c$ quarks (left panel)
and between $c$ and $\bar c$ (right panel).
}
 \label{fig:dsig_dphi}
\end{figure}

From the DPS studies point of view, the azimuthal angle correlations between
$c$ and $c$ or $c$ and $\bar c$ are very interesting. 
The corresponding distributions are shown in Fig.~\ref{fig:dsig_dphi}.
We note much bigger decorrelation of two $c$ quarks or $c$ and $\bar c$ 
in the $k_t$-factorization approach compared to the collinear approach.
This is due to explict account of gluon virtualities (transverse momenta).
We will return to this point when discussing azimuthal correlations 
between mesons at the end of this section.

\begin{figure}[!h]
\begin{minipage}{0.47\textwidth}
 \centerline{\includegraphics[width=1.0\textwidth]{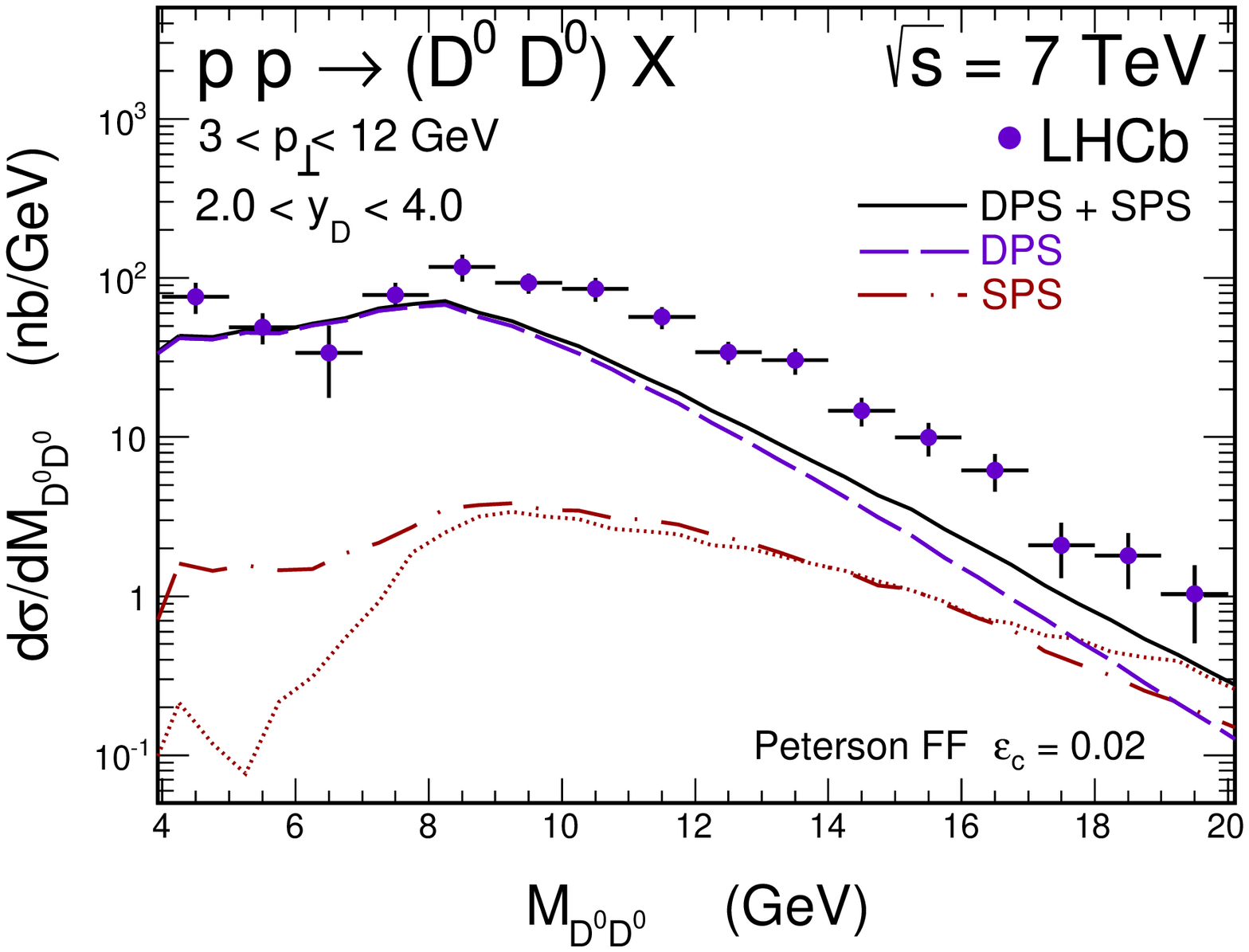}}
\end{minipage}
\hspace{0.5cm}
\begin{minipage}{0.47\textwidth}
 \centerline{\includegraphics[width=1.0\textwidth]{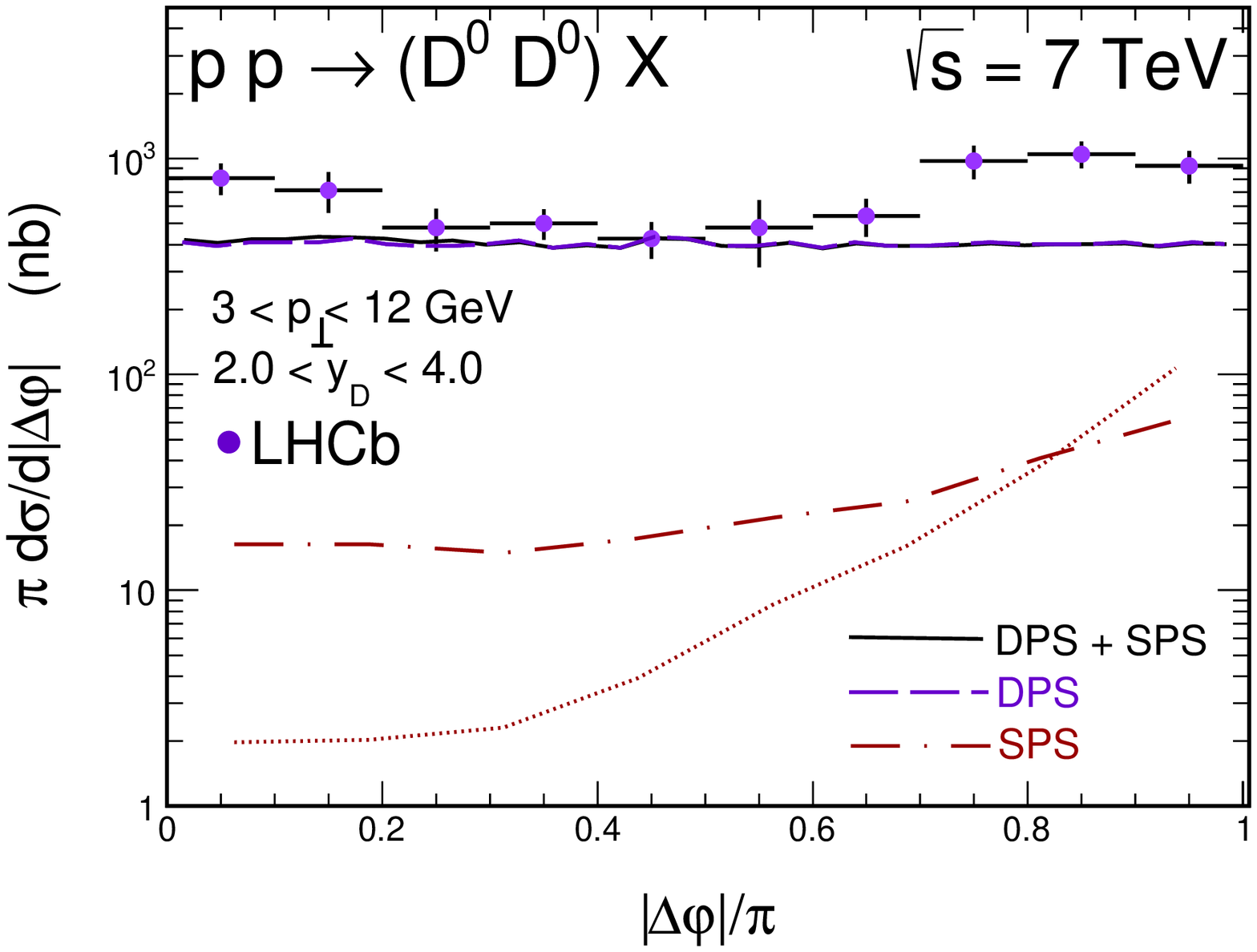}}
\end{minipage}
\caption{
\small Distributions in $D^0D^0$ invariant mass (left) and in azimuthal 
angle between both $D^0$'s (right) within the LHCb acceptance. 
The DPS contribution (dashed line) is compared with the SPS one 
calculated within the $k_t$-factorization approach (dashed-dotted line).
The SPS result from our previous studies \cite{vanHameren:2014ava}, 
calculated in the LO collinear-factorization approach, 
is also shown here (dotted line). 
}
 \label{fig:LHCb_D0D0}
\end{figure}

So far we have considered production of $c \bar c c \bar c$ quarks/antiquarks.
In the following we have included also $c \to D$ hadronization effects, 
which are important for the LHCb acceptance in meson transverse momentum.
Details how the hadronization of heavy quarks is done within 
the fragmentation function technique were explained e.g. 
in Ref.~\cite{Maciula:2013wg}.
Here we have used the Peterson fragmentation function with 
$\varepsilon_{c} = 0.02$. 
As explained in Ref.~\cite{vanHameren:2014ava} the DPS gives cross sections
very similar to those measured by the LHCb collaboration \cite{Aaij:2012dz}.
How important is the SPS contribution discussed in this paper,
calculated here in the $k_t$-factorization, is shown 
in Fig.~\ref{fig:LHCb_D0D0}. 
For comparison we show also SPS results calculated in
collinear-factorization approach \cite{vanHameren:2014ava}. 
The two approaches give somewhat different shapes of correlation
observables, inspite that the integrated cross sections are rather
similar as discussed already at the parton level.
Our results, so far the most advanced in the literature as far as the
SPS contribution is considered, are not able to explain discrepancy
between DPS contribution and the LHCb experimental data. 
Whether the discrepancies are due to simplifications in the treatment 
of DPS requires further studies including for example spin and 
flavour correlations.
Some works in this direction already started \cite{Mulders2015}.

\section{Conclusions}

We have presented results of a first calculation of the SPS cross section
for $p p \to c \bar c c \bar c X$ in the $k_t$-factorization approach,
This is a first $2 \to 4$ process for which
$k_t$-factorization is applied. In this calculation we have
used the Kimber-Martin-Ryskin unintegrated gluon distribution(s)
which effectively includes the dominant higher-order corrections.
The off-shell matrix element was calculated using a new technique
developed recently in Krak\'ow.

The results of the $k_t$-factorization approach were compared
with the results of the collinear-factorization approach.
In general, the $k_t$-factorization results are only slightly bigger than
those for collinear approach.
An exception is the transverse momentum distribution for transverse
momenta above 10 GeV where a sizeable enhancement has been observed.
Inclusion of gluon virtualities leads to a decorrelation in azimuthal
angle between $c$ and $c$ or $c$ and $\bar c$.

Since the cross section is in general very similar as for
the collinear-factorization approach we conclude that the
$c \bar c c \bar c$ final state at the LHC energies is dominantly 
produced by the double parton scattering as discussed in our recent
papers, and the SPS contribution, although interesting by itself, 
is rather small.
A comparison to predictions of double-parton scattering results
and recent LHCb data for azimuthal angle correlations between $D^0$ and $D^0$
or $\bar{D}^0$ and ${\bar D}^0$ mesons strongly suggests that the assumption 
of two fully independent DPS 
($g g \to c \bar c \otimes g g \to c \bar c$) may be too approximate
or some other mechanisms contribute.

\vspace{0.5cm}

{\bf Acknowledgments}\\
We are indebted to Andreas van Hameren for collaboration on some issues 
presented here.


\end{document}